\documentclass[prb, preprint]{revtex4}
\usepackage{framed}
\usepackage{graphicx}
\usepackage{dcolumn}
\usepackage{amsmath}
\usepackage{amssymb}
\usepackage{subfigure,amsmath,verbatim,moreverb,bm}
\usepackage{color}

\def\be{\begin{equation}}
\def\ee{\end{equation}}
\def\ber{\begin{eqnarray}}
\def\eer{\end{eqnarray}}

\def\rv{{\bf r}}

\def\pv{{\bf p}}

\def\kv{{\bf k}}

\def\nn{\nonumber}

\begin{document}
\title{Intrinsic spin Hall effect at asymmetric oxide interfaces: the role of transverse wave functions}
\author{Lorien H. Hayden$^1$}
\author{R. Raimondi$^2$}
\author{M. E. Flatt\'e$^3$}
\author{G. Vignale$^1$}
\affiliation{$^1$Department of Physics and Astronomy, University of Missouri, Columbia, MO, USA} 
\affiliation{$^2$Department of Mathematics and Physics, Universit\`a  Roma Tre, Roma, Italy}
\affiliation{$^3$Department of Physics, University of Iowa, Iowa City, USA}

\begin{abstract}
{An asymmetric triangular potential well provides  the simplest model
for the confinement of mobile electrons at the interface between two insulating oxides, such as 
 LaAlO$_3$ and SrTiO$_3$ (LAO/STO).  These electrons  have been recently shown to exhibit a large spin-orbit coupling of the Rashba type, i.e., linear in the in-plane momentum.  In this paper we study the intrinsic spin Hall effect due to Rashba coupling in an asymmetric triangular potential well.    This is the minimal model that captures the asymmetry of the spin-orbit coupling on opposite sides of the interface.  Besides splitting each subband into two branches of opposite chirality, the spin-orbit interaction causes the transverse wave function (i.e., the wave function in the $z$ direction, perpendicular to the plane of the quantum well) to depend on the in-plane wave vector $\kv$.   At variance with  the standard Rashba model, the triangular well  supports a non-vanishing intrinsic spin Hall conductivity, which is proportional to the  square of the spin-orbit coupling constant and, in the limit of low carrier density, depends only on the effective mass renormalization associated with the $\kv$-dependence of the transverse wave functions.    The origin of the effects  lies in the non-vanishing matrix elements of the spin current between  subbands corresponding to different states of quantized motion perpendicular to the plane of the well.}
 \end{abstract} 
\maketitle

\section{Introduction}
The spin Hall effect\cite{Dyakonov71} has been a topic of great interest in the last decade\cite{Hirsch99, Zhang00, Murakami03, Sinova04, Handbook, WinRev, Culcer_SteadyState_PRB07, Culcer_Generation_PRL07,Culcer_SideJump_PRB10, TsePRB05, Malshukov_Edge_PRL05, GalitskiPRB06,Tanaka_NJP09,Hankiewicz09,TenYears2010}, and it has now become a mainstream technique for the manipulation of spins in spintronic devices\cite{Ralph_MgnSwc_SHE_11,Wu_SpinDyn_PhysRep10, Fabian_Slov07, Awschalom07, Zutic04}.   Its signature is the appearance of a  current of $z$-spin in the $y$-direction following the application of  an electric field  in the $x$ direction.\cite{Kato04, Sih05, Wunderlich05, Stern06, Stern08,Ralph_Pt_SHE_Rvw_11}  The inverse effect, i.e., the generation of a transverse electric field by an injected spin current has also been observed.\cite{Valenzuela_Nat06,Takahashi_Revese_PRL07,Takahashi_GSH_NatMater08} 

It is by now clear that the spin Hall effect results from an intricate competition of several mechanisms.\cite{Nozieres73}  
In all cases one needs a ``sink of momentum", usually provided by impurities or phonons, in order to attain a steady-state response  to the applied electric field.   The spin Hall effect in a crystalline solid can be described as {\it extrinsic} or {\it intrinsic} depending on whether it is driven by  spin-orbit interaction with impurities  (extrinsic case)  or with the atomic cores  of the regular lattice (intrinsic case).  Early studies focused on the analytically solvable model of a two-dimensional electron gas in a wedge-shaped quantum well with Rashba spin-orbit coupling.\cite{Rashba84} This model is relevant to (001) GaAs quantum wells in the presence of an electric field perpendicular to the plane of the electrons.   In this model, the intrinsic and extrinsic components of the effect are distinguished by symmetry --  the former being even under a reversal of the sign of the spin-orbit coupling constant, while the latter is odd.  It was soon realized that the intrinsic spin Hall conductivity for this model vanishes exactly,\cite{Mishchenko2004,Raimondi05,Khaetskii2006} and the extrinsic spin Hall conductivity is  suppressed when the Dyakonov-Perel spin relaxation rate becomes comparable to or exceeds the Elliott-Yafet spin relaxation rate.\cite{Hankiwicz_PhaseDiag_PRL08,Raimondi09,Raimondi_AnnPhys12}  On the other hand, the vanishing of the spin Hall conductivity has been  recognized to be a peculiarity of this model.\cite{Raimondi05}  Intrinsic spin Hall conductivity has been predicted and observed in two-dimensional hole gases, and, most remarkably, in three dimensional centro-symmetric d-band metals (e.g. Pt).\cite{Takahashi_Revese_PRL07}


Recently,  a high mobility two-dimensional electron gas (2DEG) of tunable density has been observed at the interface between two insulating oxides, such as LaAlO$_3$ and SrTiO$_3$.\cite{Ohtomo04,Thiel06,Huijben06,Dagotto2007,Caviglia10}   A large spin-orbit splitting of the Rashba form, i.e., $\Delta E(\kv) = \hbar \alpha k$, where $\kv$ is the wave vector in the interfacial plane and $\hbar \alpha$ can be as large as $5 \times 10^{-2}$ eV \AA, has been observed in the proximity of a superconducting transition.\cite{Caviglia10}  In view of this large spin-orbit splitting, the interfacial 2DEG seems an excellent candidate for the observation of the spin Hall effect.  However, one must be wary of the fact that vertex corrections tend to suppress the spin Hall conductivity of systems with linear-in-$k$ spin-orbit splitting.   

In this paper, we introduce a very simple, analytically solvable  model, which we hope will help clarify the essential ingredients of the intrinsic spin Hall effect at oxide interfaces.
The model is inspired by the earlier work by Popovic and Satpathy~\cite{Popovic05}, which modeled the potential that binds the electrons to the interface as a symmetric triangular quantum well.  We generalize their model in two ways:  first, we allow for different potential slopes (i.e. electric fields) on opposite sides of the interface; second, we include a spin-orbit interaction of the Rashba type, but let it act only in the right half ($z>0$) of the well.
Thus, the Hamiltonian is
\be\label{H1}
H=\sum \left\{\frac{p^2}{2m} -\frac{\hbar^2}{2m}\frac{d^2}{dz^2} + V(z) - \frac{\lambda_c^2}{\hbar}V'(z) \Theta(z) [p_x \sigma_y-p_y \sigma_x]\right\}
\ee
where $\Theta(z)$ is the Heaviside step function ($\Theta(z)=1$ for $z>0$ and $\Theta(z)=0$ otherwise);  $\pv=(p_x,p_y)$ is the momentum in the plane of the quantum well;  $z$ is the coordinate perpendicular to the plane, and $\lambda_c$ is the effective ``Compton wavelength", which controls the strength of the spin-orbit coupling in the relevant conduction band of the quantum well.  In semiconductors like GaAs $\lambda_c$ is known to scale inversely to the cube of the fundamental band gap and amounts to a few Angstroms.  In oxide materials it is  somewhat smaller: $\lambda_c \simeq  0.7$~\AA as can be inferred from  the observed value of $\hbar \alpha = \lambda_c^2 eE \simeq 5 \times 10^{-2}$ eV \AA assuming an electric field of the order of $1$ V/\AA.  The model potential is
\be
V(z)=\left \{ \begin{array}{cc}~Fz~ & z>0\, \\
-rFz~ & z<0\, \end{array}\right.,
\ee
where $r>1$ is our asymmetry parameter and  $F=eE$, $E$ being the magnitude of the electric field in the $z$ direction and $e$ the absolute value of the electron charge.  

In spite of its simplicity, this model gives a reasonable description of electrons bound at the interface of two insulating oxides, such as SrTiO$_3$ and  LaAlO$_3$.\cite{Ohtomo04,Thiel06,Huijben06,Popovic05,Dagotto2007,Caviglia10,Ismail-Beigi2010}  We have in mind an $n$-type interface, which is equivalent to a sheet of positive charge  at $z=0$.  According to band-structure calculations\cite{Ismail-Beigi2010} the electrons that neutralize this sheet of positive charge reside primarily on the SrTiO$_3$ side ($z>0$) where both the band gap and the electric field are smaller.  On the LaAlO$_3$ side   ($z<0$)  the electric field is larger, due to reduced electrostatic screening, and this is the effect we try to capture with the parameter $r>1$. Further, the spin-orbit interaction within the conduction band is largely determined by the spin-orbit interaction of the ``B'' ion within the perovskite formula ABO$_3$; the spin-orbit interaction of the Al orbitals is negligible compared with that of the Ti orbitals, so the Rashba spin-orbit coupling will be much smaller on the LaAlO$_3$ side than on the SrTiO$_3$ side.  This is the situation modeled with the Heaviside function;  the spin-orbit coupling is significant only for $z>0$.
 All things considered, our model is probably the minimal model that captures a most significant feature of the system under study, namely the asymmetry (or, more generally, the $z$-dependence) of the spin-orbit coupling.   This feature has recently caught the attention of other researchers~\cite{Wang2013} as a possible source of novel effects at insulator-metal-insulator interfaces.  
Furthermore, a density-dependent, and hence $z$-dependent, Rashba spin-orbit coupling has also been suggested \cite{Caprara12} to explain charge inhomogeneities in LaAlO$_3$/SrTiO$_3$ systems.
Here we show that this asymmetry is entirely responsible for the appearance of $\kv$-dependent transverse wave functions and hence the non-vanishing of the intrinsic spin Hall conductivity.  On the other hand,  more realistic models for  the conduction d-electrons at the surface of SrTiO$_3$ have recently appeared in the literature,\cite{Khalsa11,Khalsa12,Khalsa13} which hold great promise to explain the transport properties of oxide interfaces.  None of these models however, seems to address the $z$-dependence of the Rashba coupling and the ensuing  
$\kv$-dependence of the transverse wave functions, which are the focal points of this paper.



The eigenfunctions of the hamiltonian~(\ref{H1})  have the form 
 \be\label{PSI}
 \psi_{n\kv\lambda}(\rv,z) = \frac{e ^{i\kv \cdot \rv}}{\sqrt{\cal A}} \frac{1}{\sqrt{2}}\left(
 \begin{array}{c} 1\\i \lambda e^{i\theta_\kv} \end{array}\right) f_{n\kv\lambda}(z)
 \ee
 where ${\cal A}$ is the area of the interface, $\kv=(k_x,k_y)$ is the in-plane wave vector, 
 $\rv$ is the position in the interfacial plane and $z$ is the coordinate perpendicular to the plane.  $\theta_\kv$ is the angle between  $\kv$ and the $x$ axis.   These states are classified by a {\it subband index} $n=0,1,2..$, which plays the role of principal quantum number,  an in-plane wave vector $\kv$, and a {\it helicity index},   $\lambda=+1$ or $-1$ which determines the form of the spin-dependent part of the wave function.   
 

The most interesting feature of this model is that the subband wave functions $f_{n\kv\lambda}(z)$  depend on $\kv$.  This feature is crucial to the existence of a non-vanishing intrinsic spin Hall conductivity.   This can be seen most clearly by applying to the present model the standard argument for the vanishing of the intrinsic spin Hall conductivity in the  Rashba model.\cite{Dimitrova05}  According to Eq.~(\ref{H1}) the time derivative of $\sigma_y$ is
\be
\dot \sigma_y = \frac{i}{\hbar}[H,\sigma_y]=-2\frac{\lambda_c^2}{\hbar^2}V'(z)\Theta(z) p_y\sigma_z\,.
\ee
The expectation value of a time derivative must vanish in a steady state, hence
\be
\langle V'(z)\Theta(z) p_y\sigma_z \rangle \equiv \sum_{n\kv\lambda}N_{n\kv\lambda} \langle n\kv\lambda|V'(z)\Theta(z) k_y\sigma_z|n\kv\lambda\rangle=0\,,
\ee
where $N_{n\kv\lambda}$ is the average occupation numbers of  $|n\kv\lambda\rangle$ in the given non-equilibrium state.
If at this point we were allowed to factor the average into a product of $\langle V'(z)\Theta(z)\rangle$ and $\langle p_y  \sigma_z\rangle$, we could immediately conclude that the spin Hall current, being proportional to $\langle p_y  \sigma_z\rangle$, is zero.  This argument works in the Rashba model because the $z$-dependent part of the wave function does not depend on $\kv$ or $\lambda$.  In the present case, however, the $z$-dependent wave functions $f_{n\kv\lambda}(z)$ do depend on $\kv$ and $\lambda$, creating a correlation between $V'(z)\Theta(z)$ and  $p_y\sigma_z$.  Then, we can no longer assert that the spin Hall current vanishes.  In fact, writing 
\be
\langle V'(z)\Theta(z) p_y\sigma_z \rangle = \langle V'(z)\Theta(z)\rangle \langle p_y  \sigma_z\rangle +\langle \Delta[V'(z)\Theta(z)]\Delta[p_y\sigma_z] \rangle 
\ee
where $\Delta[A]$ represents the fluctuation of a quantity relative to its average, we can conclude that
\be
 \langle p_y\sigma_z \rangle = - \frac{\langle \Delta[V'(z)\Theta(z)]\Delta[p_y\sigma_z] \rangle}{\langle V'(z)\Theta(z)\rangle}
 \ee  

Our calculations confirm this.   The shape of the confining potential in our model is controlled by the asymmetry parameter $r$.   For $r \to \infty$ the electrons are entirely confined to the right half of the quantum well.
In this limit we recover the  Rashba model. The subband wave functions become independent of $k$, because the spin-orbit coupling is independent of $z$ in the region of space in which the electrons move. The spin Hall conductivity vanishes, when  the ``vertex correction" is duly taken into account.   

For finite $r$ additional contributions of order $\alpha^2$ appear from the $k$-dependence of the transverse wave functions.   Here $\alpha$ is the dimensionless spin-orbit coupling constant 
\be
\alpha\equiv \left(\frac{\lambda_c}{\ell}\,\right)^2,
\ee
where 
\be\label{LengthScale}
\ell \equiv \left(\frac{\hbar^2}{2m F}\right)^{1/3}
\ee
is the natural length scale of our model.  Since $\ell$ is of the order of  a few Angstroms for oxide interfaces, while $\lambda_c \simeq 1$\AA, we see that $\alpha$ is somewhat smaller than $1$, but not orders of magnitude smaller.  We believe that a multi-band effect, proportional to $\alpha^2$, is also responsible for the intrinsic spin Hall conductivity predicted and observed  in centro-symmetric metals like Pt.\cite{Tanaka_NJP09}   Indeed, we have verified that the $\alpha^2$ spin Hall conductivity is present in our model even if we make the spin-orbit coupling symmetric and set $r=1$, thus simulating a centro-symmetric material.  In this case the subbands become  doubly degenerate with respect to the helicity index, but  the $\alpha^2$ contribution to the spin Hall conductivity is still present.  

Assuming, as discussed above,  that our model gives a reasonably good description of electrons at oxide interfaces, the results of our work imply that the intrinsic spin Hall effect will be an effect of order $\alpha^2$ (as opposed to $\alpha^0$) and be crucially dependent on the energy separation between the transverse subbands.

This paper is organized as follows.  In Section II we discuss the analytic solution of the model.
In Section III we calculate the inter-subband contributions to the SHC as a function of $r$.
In Section IV we calculate the intra-subband contribution to the spin Hall effect - first without including vertex corrections (where it is found to be quite large and independent of $\alpha$ ) and then with vertex corrections (the correct approach).  In the latter case, only a contribution proportional to $\alpha^2$ survives.

\section {Solution of the model}
We express all quantities in units derived from the natural length scale 
 $\ell$ defined in Eq.(\ref{LengthScale}),  i.e., lengths in units of $\ell$, momenta in units of $\hbar /\ell \simeq 1$ \AA$^{-1}$,   energies  in units of $F\ell=\frac{\hbar^2}{2m\ell^2} \simeq 1$ eV.  All quantities in the following treatment are therefore dimensionless, unless  noted otherwise.
 The dimensionless hamiltonian takes the form
 \be
H=\sum \left\{k^2 -\frac{d^2}{dz^2} + v(z) - \alpha \Theta (z) v'(z) [k_x \sigma_y-k_y \sigma_x]\right\}
\ee
where $\kv=\pv/\hbar$ in physical units, and $\kv=\pv$ in the present reduced units. 
Here we have defined
\be
v(z)=\left \{ \begin{array}{cc}~z~ & z>0\,, \\
-rz~ & z<0\,. \end{array}\right.
\ee

 We note that $k_x$, $k_y$ and $k_x \sigma_y-k_y\sigma_x$ are compatible constants of the motion, the latter with eigenvalues $\lambda k$ with $\lambda=\pm1$ and eigenstates of the form
 \be
 \frac{1}{\sqrt{2}} \left(\begin{array}{c} 1\\i \lambda e^{i\theta_\kv} \end{array}\right)\,,
\ee
where $\theta_\kv$ is the angle between $\kv$ and the $x$ axis.
 Therefore we  classify the eigenstates by quantum numbers $\kv$ (in-plane wave vector) and $\lambda = \pm 1$ (helicity).
 The wave functions for the $n$-th subband ($n=0,1,2...$ in order of increasing energy)  have the form given in Eq.~(\ref{PSI})
where
 $f_{nk\lambda}(z)$ is the solution of the Schr\"odinger equation
 \be\label{SE}
 -f''_{nk\lambda}(z)+[v(z)-\lambda \alpha \Theta(z) kv'(z)] f_{nk\lambda}(z) =  \epsilon_{n\kv \lambda}f_{nk\lambda}(z)
 \ee
  with
  \be
v'(z)=\left \{ \begin{array}{cc}~1~ & z>0\,, \\
-r~ & z<0\,. \end{array}\right.
\ee 
 The eigenvalues $\epsilon_{n\kv\lambda}$ are arranged in increasing order,
 \be
 \epsilon_{0k\lambda} <\epsilon_{1k\lambda}<\epsilon_{2k\lambda}<...
 \ee  
  The complete energy is
  \be
  E_{n\kv\lambda}= k^2+\epsilon_{n\kv\lambda}\,.
  \ee
  
  The solution of the Schr\"odinger equation~(\ref{SE}) for generic $n$ is
 \be\label{FK}
f_{k\lambda}(z) = Z\times\left\{
\begin{array}{c} \frac{Ai(z-\epsilon_{\kv\lambda}-\lambda\alpha k)}{Ai(-\epsilon_{k\lambda}-\lambda\alpha k)}~~~~(z \geq0)\\
 \frac{Ai(-zr^{1/3}-\epsilon_{k\lambda}r^{-2/3})}{Ai(-\epsilon_{\kv\lambda}r^{-2/3})} ~~~~(z <0)
\end{array}
\right.\,,
\ee 
where $Ai(x)$ is the Airy function and  $Z$ is the normalization constant.  Notice that this wave function is continuous at $z=0$, with $f_{\kv\lambda}(0)=Z$.
The energy $\epsilon_{\kv\lambda}$ is determined by imposing the continuity of the derivative at $z=0$:
\be
\frac{Ai^\prime(-\epsilon_{k\lambda}-\lambda\alpha k)}{Ai(-\epsilon_{k\lambda}-\lambda\alpha k)}=-r^{1/3}\frac{Ai^\prime(-\epsilon_{k\lambda}r^{-2/3})}{Ai(-\epsilon_{k\lambda}r^{-2/3})}\,,
\ee
where the $Ai^\prime(z) \equiv \frac{d Ai(z)}{dz}$.

\begin{figure}
\begin{center}
\includegraphics[width=4.0in]{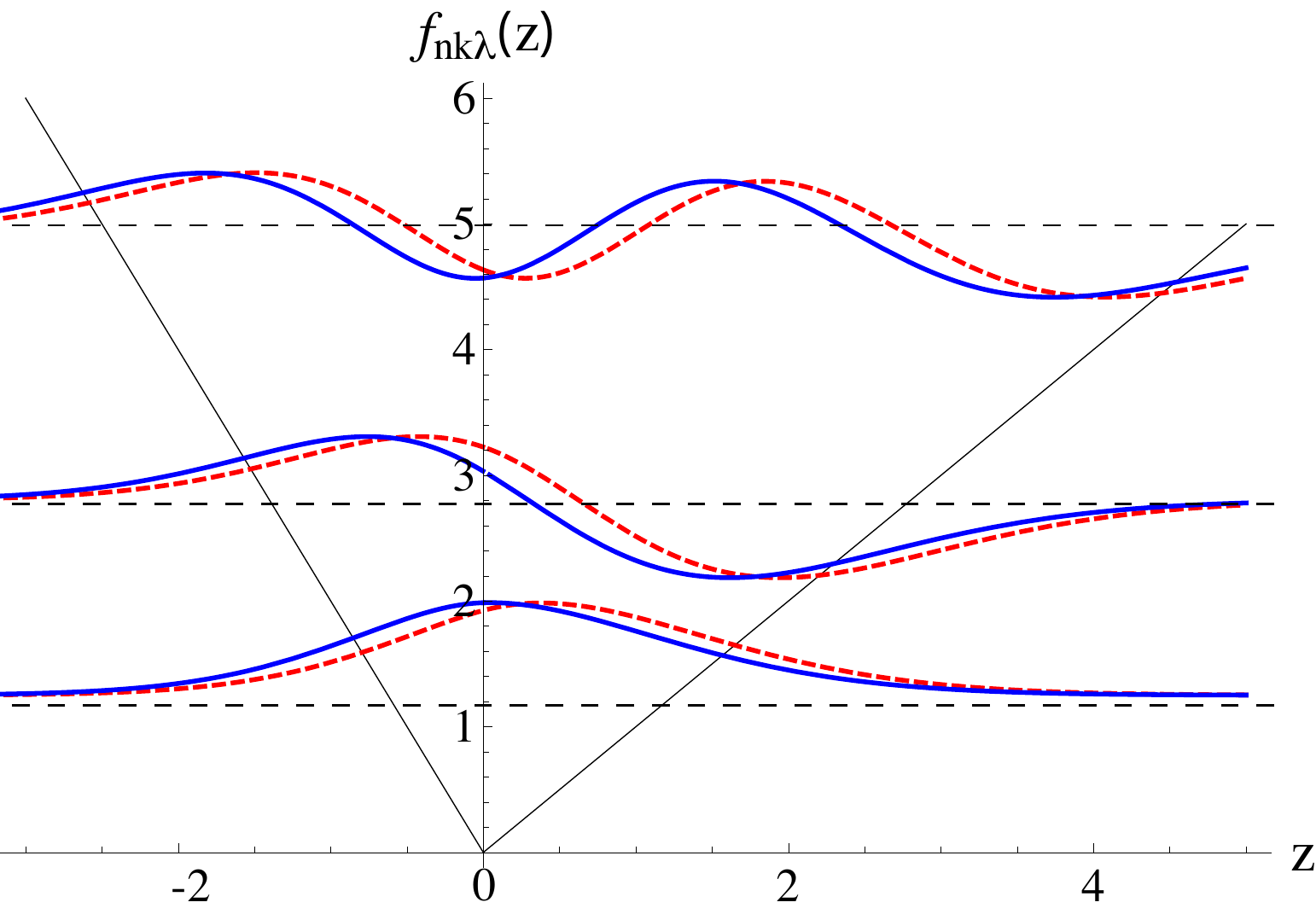}
\end{center}
\caption{Wave functions for the three lowest subbands plotted vs $z$ for $\alpha k =0.2$ for  $r=2$.  Notice the difference between $\lambda=1$ and $\lambda=-1$ wavefunctions with the same $n$, denoted by dashed (red) and solid (blue) lines respectively.  Notice that the vertical axis has been shifted to  separate clearly the wave functions with different $n$.  In reality, all wave functions tend to $0$ for $|z| \to \infty$.}
\label{Fig1}
\end{figure}

For small $k$ the subband energies and wave functions can be obtained from standard  perturbation theory.  Let us denote by $f_n(z)$ the solutions of Eq.~(\ref{SE}) at $k=0$ and by $\epsilon_n$ the corresponding energies (clearly, these are independent of $\alpha$ and $\lambda$).  Then, we have
 \be\label{f-expansion}
 f_{nk\lambda}(z)= f_n(z)+\lambda\alpha k \sum_{n' \neq n} \frac{p_{n'n}}{\epsilon_n-\epsilon_{n'}}  f_{n'}(z)+...\,,
 \ee
 where
\be
p_{n'n}\equiv \langle f_{n'}|v'(z)\Theta(z)|f_n\rangle =  \int_0^\infty f_{n'}(z)  f_n(z)dz\,, 
\ee
and
%
 \be
 \epsilon_{nk\lambda}= \epsilon_{n0}- \lambda \alpha k p_{nn} + (\alpha k)^2 \sum_{n' \neq n} \frac{|p_{n'n}|^2}{\epsilon_n-\epsilon_{n'}}+ ...~ \,.
 \ee
Bands of equal $n$ and opposite helicities cross at $k=0$.  The intra-subband splitting is linear in $k$ and proportional to $p_{nn}$.   The quadratic correction indicates  the emergence of a spin-orbit-induced effective mass $m^*$, which is related to the bare band  mass $m$ by
\be\label{EffectiveMass1}
\frac{m}{m^*}=1+ \alpha^2 \sum_{n' \neq n} \frac{|p_{n'n}|^2}{\epsilon_n-\epsilon_{n'}}\,.
\ee

All  of the above features are  confirmed by  numerical calculations (see figure below).  

  
\begin{figure}
\begin{center}
\includegraphics[width=3.0in]{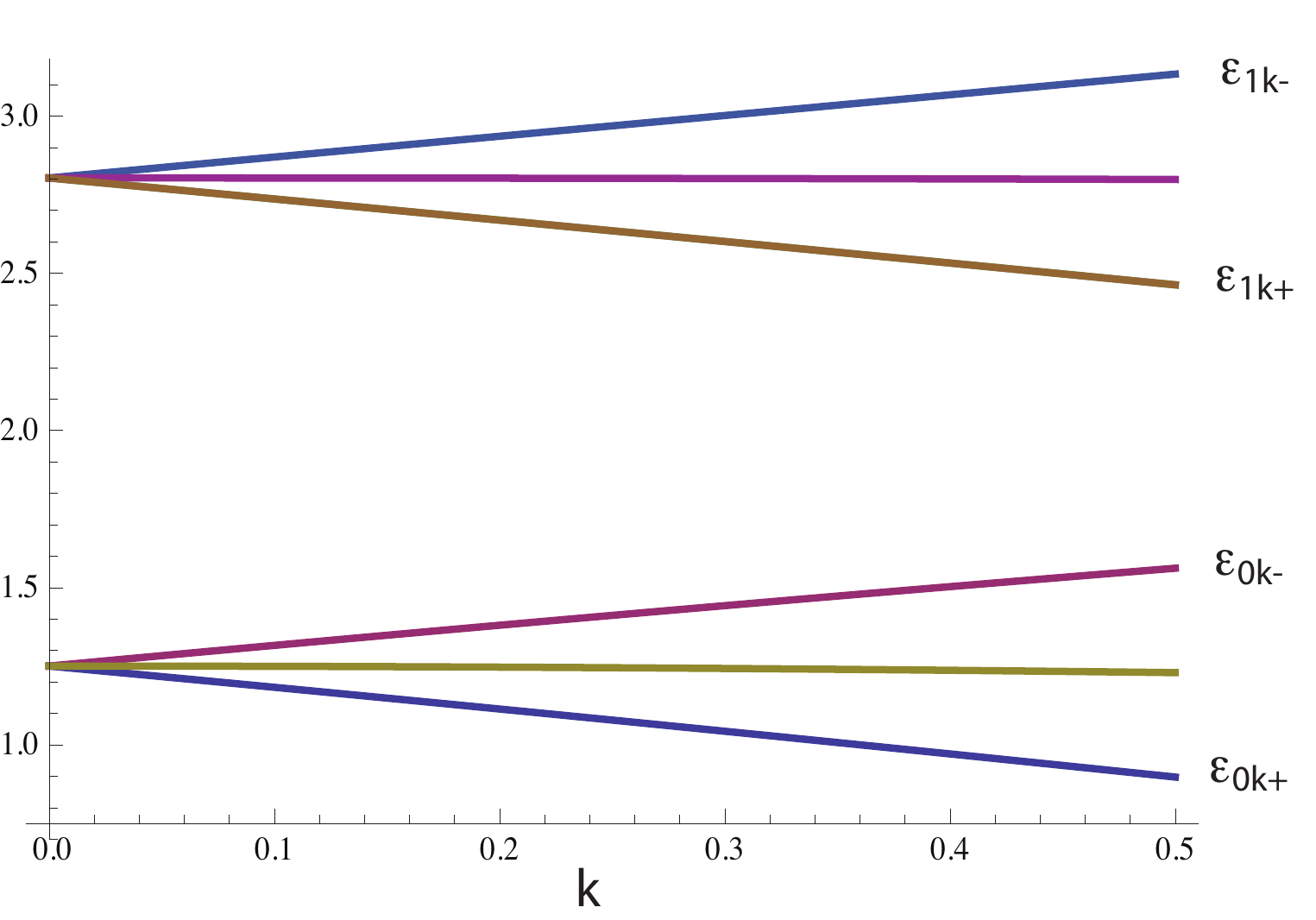}
\end{center}
\caption{The two lowest subband energies $\epsilon_{n\kv\lambda}$ (in units of $\frac{\hbar^2}{2 m\ell^2}$)  plotted as functions of $\alpha k$ for  $r=2$.  Notice the linear splitting between opposite helicities. The middle line in each case is the average of the two helicity bands and shows the (barely perceptible) curvature associated with the effective mass renormalization.}
\label{Fig2}
\end{figure}

 The following facts are emphasized, which will play a key role in the calculation of the spin Hall conductivity:
\begin{enumerate}
\item The subband wave functions are functions of the product $\alpha k$.   Wave functions of opposite helicities $f_{nk+}(z)$ and $f_{nk-}(z)$ are generally different  (see Fig.~\ref{Fig1}).  This difference disappears only for $\alpha k=0$ (i.e., either $k=0$ or $\alpha =0$).  This situation is in stark contrast with the conventional Rashba model, in which the subband wave functions are independent of $\alpha$, $k$ or $\lambda$.

\item Subband wave functions of the same $k$ and {\it opposite helicities} are not mutually orthogonal, even when their subband indices are different.  From Eq.~(\ref{f-expansion}) we see that
\be\label{Overlap-nn'}
\langle f_{nk\lambda}|f_{n'k-\lambda}\rangle= 
-2\alpha k \lambda \frac{p_{n'n}}{\epsilon_n-\epsilon_{n'}}\,,~~~~~~n' \neq n\,. 
\ee
up to terms of order $(\alpha k)^2$. 
On the other hand, for  $n=n'$ we have 
\ber\label{Overlap-nn}
\langle f_{nk\lambda}|f_{nk\lambda}\rangle 
&=& 1\nn\\
\langle f_{nk\lambda}|f_{nk -\lambda}\rangle 
&=& 1-2(\alpha k)^2\sum_{n'\neq n} \left\vert\frac{p_{n'n}}{\epsilon_n-\epsilon_{n'}}\right\vert^2\,.
\eer
To derive these overlaps it is essential to take into account second-order corrections to $f_{nk\lambda}$, ensuring that the latter remains normalized to $1$ to second order in $\alpha k$.
More generally, we have
\ber\label{Overlap-nn2}
\langle f_{nk\lambda}|f_{nk'\lambda}\rangle 
&=& 1-\frac{1}{2}[\alpha(k-k')]^2\sum_{n'\neq n} \left\vert\frac{p_{n'n}}{\epsilon_n-\epsilon_{n'}}\right\vert^2\nn\\
\langle f_{nk\lambda}|f_{nk' -\lambda}\rangle 
&=& 1-\frac{1}{2}[\alpha(k+k')]^2\sum_{n'\neq n} \left\vert\frac{p_{n'n}}{\epsilon_n-\epsilon_{n'}}\right\vert^2\,.
\eer
These formulas will be needed later in the analysis of vertex corrections.
\end{enumerate}

\section{Intrinsic spin Hall effect}
In this section we begin  the calculation of the intrinsic spin Hall conductivity $[\sigma_{SH}]^z_{yx}$ which connects the $j^z_y\equiv k_y\sigma_z$ component of the spin current  to an electric field applied in the $x$ direction.
The real part of the spin Hall conductivity is given by the formula
	\be\label{sigmaSH-1particle}
	[\sigma_{SH}]^z_{yx}=-\frac{e}{{\cal A}}\sum_{n\kv\lambda,~occ} \Im m  \left\langle\left.\frac{\partial \psi_{n\kv\lambda}}{\partial 
	A^z_y}\right\vert \frac{\partial \psi_{n\kv\lambda}}{\partial A^0_x}\right \rangle\,,
	\ee
where the sum runs over the occupied states.  We will assume in what follows that only states in the lowest subband (i.e., the subband with $n=0$) are significantly populated in the ground state.  We use perturbation theory to calculate the variation of the wave functions $\psi_{n\kv\lambda}$ to the application of infinitesimal vector potentials that couple to $k_x$ and $k_y$ respectively, the first shifting $k_x \to k_x+A^0_x$,  the second shifting $k_y \to k_y+A^z_y\sigma_z$ (the hamiltonian must be properly symmetrized after the second replacement, to preserve hermiticity).  It is evident that these homogeneous perturbations do not mix wave functions of different $\kv$. For the asymmetric well, it is not necessary to use degenerate perturbation theory. Instead, the usual non-degenerate theory suffices. The zeroth order wavefunctions are given by Eq.~(\ref{PSI}), where the subband wave functions $f_{nk\lambda}(z)$ are given by Eq.~(\ref{FK}), evaluated at the appropriate values of the energy.
In order to apply our formula (\ref{sigmaSH-1particle}) we need to find the single-particle wave functions to first order in the applied potentials $A_x^0, A_y^z$. We note that the application of $A_x^0$ modifies the wavefunction at a given $\mathbf{k},\lambda$ in the following manner:
	\begin{equation}
	\psi_{n\mathbf{k}\lambda}[A^0_x](\mathbf{r})=\frac{1}{\sqrt{2}}\begin{pmatrix}1\\  i\lambda 
	e^{i\theta_{\tilde\kv}}\end{pmatrix}f_{n\tilde\kv\lambda}(z) e^{i\mathbf{k}\cdot\mathbf{r}}\,,
	\end{equation}
where $\tilde\kv=\kv+A_x^0\hat{\bf x}$ is the shifted wave vector. 
Thus, specializing to the $n=0$ subband, we find
	\begin{equation}\label{Derivative1}
	\frac{\partial\psi_{0\mathbf{k}\lambda}(\mathbf{r})}{\partial A_x^0}\bigg{|}_{A_x^0=0}=
\lambda\frac{\sin\theta_\mathbf{k}}	
	{\sqrt{2} k}\begin{pmatrix}0\\  
	e^{i\theta_{\mathbf{k}}}\end{pmatrix}f_{0\kv\lambda}(z)  e^{i\kv\cdot\rv}+\frac{\cos\theta_\kv}{\sqrt{2}}\begin{pmatrix}1\\  
	i\lambda e^{i\theta_{\kv}}\end{pmatrix}\frac{\partial f_{0 k\lambda}(z)}{\partial k}e^{i\kv\cdot\rv}.
	\end{equation}
Next, we want to calculate $\left.\frac{\partial \psi_{0\kv\lambda}(\rv)}{\partial A_y^z}\right\vert_{A_y^z=0}$ To do this, we observe that the first  order change in the wavefunction due to the coupling to $A_y^z$ is
	\begin{equation}
	\delta\psi_{0\kv\lambda}(\rv)=2A_y^z k_y \sum_{n}\frac{\langle f_{n\kv\bar\lambda}|f_{0\kv\lambda}\rangle}{\epsilon_{0\kv\lambda}-\epsilon_{n\kv\bar\lambda}}\psi_{n\kv\bar\lambda}(\rv)+...\,,
	\end{equation}
where $\bar \lambda$ is a short-hand for $-\lambda$.  The reason why only states of helicity $-\lambda$ appear in the variation of $\psi_{0\kv\lambda}$ is that the $A_y^z$ field couples to the spin current operator $k_y\sigma_z$, which flips the helicity index.
Writing $k_y$ as $k \sin\theta_{\mathbf{k}}$ yields
	\begin{equation}\label{Derivative2}
	\left.\frac{\partial \psi_{0\kv\lambda}(\rv)}{\partial A_y^z}\right\vert_{A_y^z=0}= 2 k \sin\theta_{\mathbf{k}}\sum_{n}\frac{\langle f_{n\kv\bar\lambda}|f_{0\kv\lambda}\rangle}{\epsilon_{0\kv\lambda}-\epsilon_{n\kv\bar\lambda}}\psi_{n\kv\bar\lambda}(\rv)
	\end{equation}
Lastly, we combine Eqs.~(\ref{Derivative1}) and~(\ref{Derivative2}) for the derivatives of the wave function into the general formula~(\ref{sigmaSH-1particle}) for the spin Hall conductivity.  Only the first term on the right hand side of Eq.~(\ref{Derivative1})  contributes, due to the vanishing of the angular average of $\sin\theta_\kv \cos\theta_\kv$.  Taking into account the fact that the lowest subband states are populated in the ground state up to $k=k_{F+}$ for helicity $+1$ and up to $k=k_{F-}$ for helicity $-1$   we arrive, after some simple algebra, to the following expression, 
	\begin{equation}\label{SHC2}
	\begin{split}
	[\sigma_{SH}]_{yx}^z=
	-\frac{e}{4\pi}\sum_{n}\bigg{[}\int_{0}^{k_{F+}}\frac{|\langle f_{n\kv-}|f_{0\kv+}\rangle|^2}{\epsilon_{0\kv+}- \epsilon_{n\kv-}}k 
	dk+\int_0^{k_{F-}}\frac{|\langle f_{n\kv+}|f_{0\kv-}\rangle|^2}{\epsilon_{0\kv-}-\epsilon_{n\kv+}}kdk\bigg{]}
	\end{split}
	\end{equation}
	All the quantities that appear in the square brackets of this equation are dimensionless.  We now examine separately the inter-subband contributions ($n \neq 0$) and the intra-subband contributions ($n=0$) to the spin Hall conductivity.

\subsection{Inter-subband contribution}

\begin{figure}
	\begin{center}
	\includegraphics[width=3.0in]{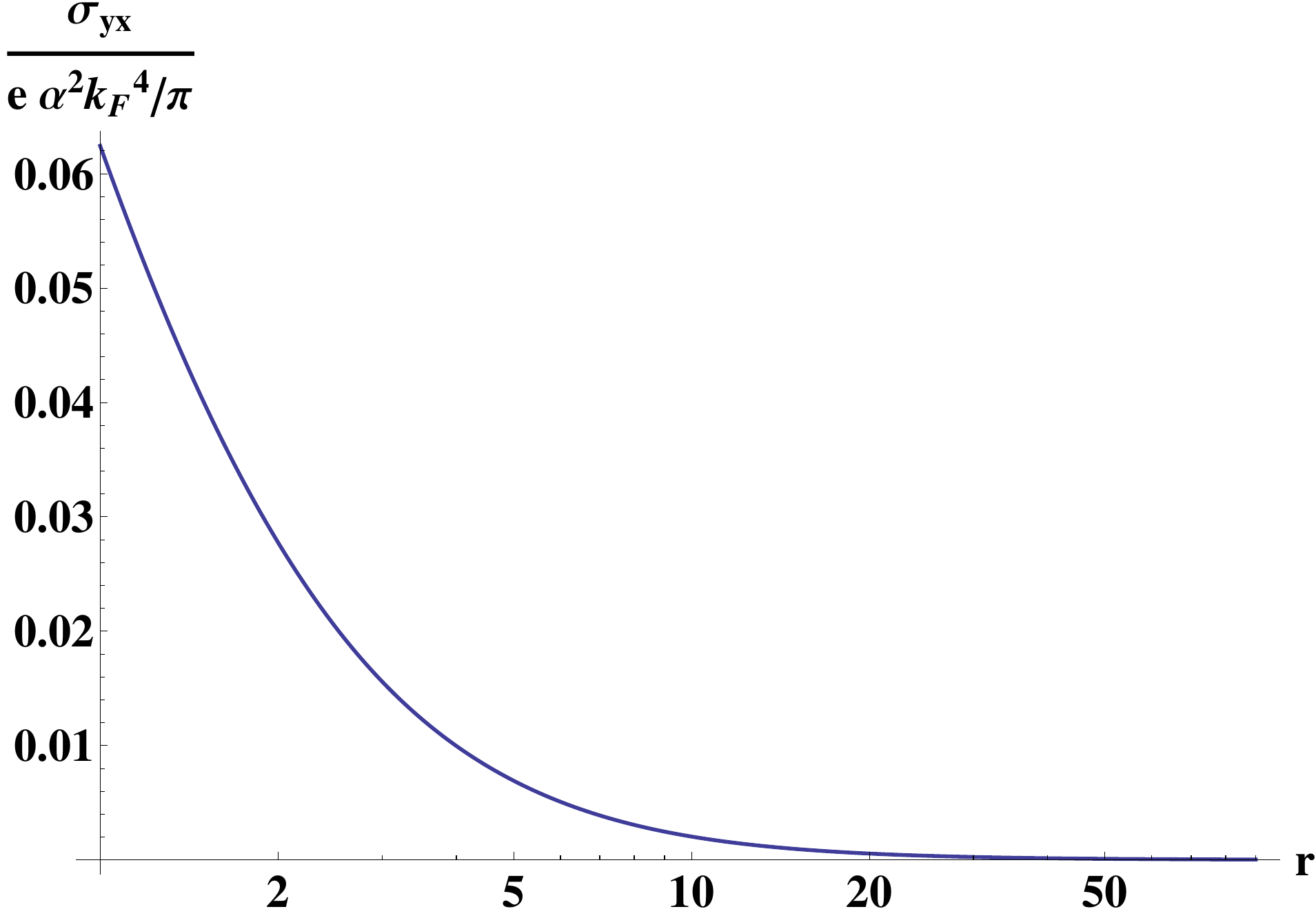}
	\caption{Intersubband contribution to  the spin Hall conductivity in units of $\frac{e \alpha^2 k_F^4}{\pi}$ (see Eq.~(\ref{InterSubband})).   Data points have been connected to enhance visibility of the trend.  Contributions up to $n=5$ have been included.  Notice the logarithmic scale for $r$.} 
	\end{center}
	\label{Fig3}
	\end{figure}

Let us first focus on the contribution of the terms with $n \neq 0$. It is safe to assume that $k_{F+}$ and $k_{F-}$ are sufficiently small to justify the use of the small-$k$ expression (\ref{Overlap-nn'}) for the overlap between wave functions of opposite helicity.  In this approximation we can also ignore the difference between $k_{F+}$ and $k_{F-}$, i.e., we set
$k_{F+}= k_{F-}=k_F$.  Further, we can ignore the $k$-dependence of the subband energies, i.e., we set $\epsilon_{n\kv\lambda} \simeq \epsilon_{n}$. 
Then, making use of Eq.~(\ref{Overlap-nn'}) and carrying out the $k$ integral, we arrive at:
	\begin{equation}\label{InterSubband}
	[\sigma_{SH}^{\rm inter}]_{yx}^z = -\frac{e}{\pi}\alpha^2k_F^4\sum_{n\neq 0}\frac{|p_{no}|^2}{(\epsilon_0-\epsilon_n)^3}\,.
	\end{equation}
\indent
All the quantities on the right hand side, except $e$, are dimensionless. {Thus $k_F=\sqrt{2\pi n\ell^2}$, where $n$ is the two-dimensional electronic density and $\ell$ is the length scale of the confining potential.  For $\ell=3$ \AA~ and $n= 10^{14}$ cm$^{-2}$ we have $k_F\simeq 0.75$.} 
Although derived for the clean limit -- i.e., for frequencies that are  much larger than the inverse of the electron impurity scattering time $1/\tau$ (while still much smaller than the Fermi energy $E_F$)  --  this result is essentially exact for the d.c. transport regime as well.   Vertex corrections do not modify it significantly, since they are of the order $(\Delta\tau)^{-1}\ll 1$ where $\tau$ is the momentum relaxation time and $\Delta$ is the energy separation between the subbands.

 The numerically calculated inter-subband contribution to the spin Hall conductivity is plotted  ({in units of $\frac{e\alpha^2k_F^4}{\pi}$)}  in Fig.~\ref{Fig3}. We note that it is present for all values of $r$, but it vanishes in the Rashba limit ($r \to \infty$)  because $p_{n0}$ vanishes in that limit.  
 We also notice that this contribution to the spin Hall conductivity vanishes in the limit of low carrier density, i.e., for $k_F \to 0$.
	
\subsection{Intra-subband contribution}
Let us now consider the intra-subband contribution
	\begin{equation}
	 [\sigma_{SH}^{\rm intra}]^z_{yx}=-\frac{e}{4\pi}\int_{k_F-}^{k_{F+}}\frac{|\langle f_{-0}|
	f_{+0}\rangle|^2}{\epsilon_{0\kv+}-\epsilon_{0\kv-}}k~dk\,,
	\end{equation}
where we have exploited the fact that $|\langle f_{0\kv-}|f_{0\kv+}\rangle|^2=|\langle f_{0\kv+}|f_{0\kv-}\rangle|^2$ to combine the two integrals on the right hand side of Eq.~(\ref{SHC2}) into a single one.
%


The wave vectors $k_{F+}$ and $k_{F-}$ are determined by the equations
\be
k_{F+}^2+k_{F-}^2=2 k_F^2\,,
\ee
where $k_F^2=2\pi n\ell^2$, and
\be
E_{0k_{F+}+}=E_{0k_{F-}-}
\ee
We expand the energy up to third order in $\alpha$:
\be\label{EnergyExpansion}
E_{0 k\lambda}=k^2+e_1 (\alpha k) \lambda + e_2 (\alpha k)^2+e_3(\alpha k)^3\lambda
\ee
where
\begin{eqnarray}
e_1&=&-p_{00} \label{soe_12}\\
 e_2&=&-\sum_{n\neq0}\frac{|p_{0n}|^2}{\epsilon_n -\epsilon_0} =\frac{1}{\alpha^2} \left(\frac{m}{m^*}-1 \right)\label{soe_13}\\
 e_3&=&-\left[\sum_{n,m\neq0}\frac{p_{0n}p_{nm}p_{m0}}{(\epsilon_n-\epsilon_0)(\epsilon_m -\epsilon_0)} -p_{00}\sum_{n\neq0} \frac{|p_{0n}|^2}{(\epsilon_n-\epsilon_0)^2}\right] \label{soe_14}\,.
 \end{eqnarray}
 Then the solution for the Fermi wave vectors can be written in the form
  \begin{equation}
 \label{soe_2}
 k_{F\lambda}=k_F+ k_1 \alpha\lambda+k_2\alpha^2+k_3\alpha^3\lambda.
 \end{equation}
 where
 \begin{eqnarray}
 k_1&=&-\frac{e_1}{2} \label{soe_3}\\
 k_2&=&-\frac{e_1^2}{8k_F} \label{soe_4}\\
 k_3&=& \frac{e_1e_2}{2}-\frac{e_3 k_F^2}{2}.\label{soe_5}
 \end{eqnarray}

Armed with these results we easily find that
\be
k_{F+}-k_{F-}\simeq -\alpha e_1-\alpha^3k_F^2e_3+\alpha^3 e_1e_2\,.
\ee
and
\be
\epsilon_{0\kv+}-\epsilon_{0\kv-} \simeq 2\alpha k e_1+2(\alpha k)^3e_3\,.
\ee
We also know that (see Eq.~(\ref{Overlap-nn}))
\be
|\langle f_{0k+}|f_{0k-}\rangle|^2 
\simeq 1-4(\alpha k)^2\sum_{n\neq 0} \left\vert\frac{p_{n0}}{\epsilon_0-\epsilon_{n}}\right\vert^2\,.
\ee
Combining these results we  arrive at
\be\label{SHC-Intra1}
 [\sigma_{SH}^{\rm intra}]^z_{yx}=\frac{e}{8\pi} \frac{m^*}{m}\left[1- 4(\alpha k_F)^2\sum_{n\neq 0} \left\vert\frac{p_{n0}}{\epsilon_0-\epsilon_{n}}\right\vert^2\right]\,,
 \ee
 where we have used the fact that $2-\frac{m}{m^*}\simeq \frac{m^*}{m}$.
This result is striking, because its leading order is independent of both $\alpha$ and $k_F$.  However, as we now show, this result is completely invalidated by vertex corrections, which cancel the $\alpha^0$ term and leave us with an $\alpha^2$ term, which remains finite  when $k_F \to 0$.
 
\subsection{Vertex corrections}

In this subsection we discuss the vertex corrections to the  spin Hall conductivity. To this end, it is  practical to  resort to the standard diagrammatic formulation in terms of the Green functions.  The zero-temperature Kubo formula for the spin Hall conductivity reads
\begin{equation}
\label{a_1}
\left[ \sigma_{SH} \right]^z_{yx}=-e\lim_{\omega \rightarrow 0}\frac{1}{\omega}
\int_{-\infty}^{\infty}\frac{{\rm d}\epsilon}{2\pi} {\rm Tr}\Big\{{\hat \gamma}_y^z G (\epsilon_+) {\hat \gamma}_x G(\epsilon_-)  \Big\},
\end{equation}
where the trace is taken in the basis of the exact eigenstates  $|l\rangle\equiv \psi_{n\kv\lambda}(\rv,z)$.
The energies in the Green functions are $\epsilon_{\pm}=\epsilon\pm \omega /2$, and $e>0$ is the unit charge.
In the reduced units,  the current vertices read
${\hat \gamma}_y^z=k_y\sigma^z$ and $ {\hat \gamma}_x=-\alpha \Theta (z)\sigma^y$.
Notice that $v'(z)=1$ for $z>0$  has been absorbed into $\alpha$ and only the anomalous spin-dependent part of ${\hat \gamma}_x$ has been considered, since the ``regular" part does not contribute to the spin Hall conductivity.
In the  basis of the exact eigenstates the Green function is diagonal and reads
\begin{equation}
\label{a_2}
G_l(\epsilon)=\frac{1}{\epsilon -E_l +{\rm i}0^+{\rm sgn} (E_l)},   
\end{equation}
$E_l\equiv E_{n\kv \lambda}=k^2+\epsilon_{n\kv\lambda}$ being the energy eigenvalues.
By inserting the resolution of the identity, ${\hat I}=\sum_l |l\rangle\langle l|$, twice under the trace and performing the integral over $\epsilon$, we get
\begin{equation}
\label{a_3}
\left[ \sigma_{SH} \right]^z_{yx}=-e\lim_{\omega \rightarrow 0}\frac{1}{\omega}
\sum_{ll'}{\rm i}\langle l|{\hat \gamma}_y^z  |l'\rangle \langle l'| {\hat \gamma}_x|l\rangle
\frac{\Theta (-E_l)-\Theta (-E_{l'})}{\omega -E_{l'}+E_l +{\rm i}0^+}.
\end{equation}
By exploiting the fact that ${\rm i}\langle l|{\hat \gamma}_y^z  |l'\rangle \langle l'| {\hat \gamma}_x|l\rangle$ is real, after using the Kramers-Kronig
relations, we can rewrite the spin Hall conductivity in the form
\begin{equation}
\label{a_4}
\left[ \sigma_{SH} \right]^z_{yx}=e\sum_{ll'}{\rm i}\langle l|{\hat \gamma}_y^z  |l'\rangle \langle l'| {\hat \gamma}_x|l\rangle
\frac{\Theta (-E_l)-\Theta (-E_{l'})}{ (E_l -E_{l'})^2}.
\end{equation}
By making the shifts $k_x\rightarrow k_x +A_x$ and
$k_y\rightarrow k_y+A_y^z\sigma^z$ we obtain the {\it bare} current and spin-current vertices:
\ber
\nonumber
{\hat \gamma}_x&=&\frac{\partial H}{\partial A_x}\simeq -\alpha \Theta (z)\sigma^y\,,
\nn\\
{\hat \gamma}_y^z&=&\frac{1}{2}\frac{\partial H}{\partial A_y^z}=k_y\sigma^z\,,
\eer
where the $\simeq$ in the first equation reminds us that the  ``regular" part of the charge current vertex has been omitted.

The relevant matrix elements of these current vertices are the ones between states of opposite helicity:
\begin{eqnarray}
\langle n\kv \lambda | {\hat \gamma}_y^z|n'\kv \bar\lambda \rangle &=&   k_y  \langle f_{n\kv\lambda}|f_{n'\kv\bar\lambda}\rangle\label{a_4b}\\
\langle  n\kv \lambda| {\hat \gamma}_x|n'\kv \bar \lambda\rangle &\simeq& 
-{\rm i}  \frac{\sin \theta_{\kv}}{2k}(E_{n\kv\lambda} 
-E_{n'\kv\bar \lambda})\langle f_{n\kv\lambda}|f_{n'\kv\bar\lambda}\rangle.\label{a_4c}
\end{eqnarray}
Eq.(\ref{a_4b}) can be derived almost by inspection. Eq.(\ref{a_4c}) can be obtained by using the eigenvalue equation for the functions $f_{n\kv\lambda}(z)$.

Following the procedure described in Ref.~\onlinecite{Raimondi05} we calculate the renormalized current vertex $\hat \Gamma_x$ according to the equations
\ber
\label{a_7}
{\hat \Gamma}_x &=&{\tilde \gamma}_x +\frac{1}{2\pi N_0 \tau}\sum_{\kv'} G^R_{\kv'}{\hat \Gamma}_xG^A_{\kv'}\,, \nn\\
{\tilde \gamma}_x&=&-\alpha \Theta (z)\sigma^y +\frac{1}{2\pi N_0 \tau}\sum_{\kv'} G^R_{\kv'}\ 2k_x' \ G^A_{\kv'}\,.
\eer
Here $N_0=1/4\pi$ is the dimensionless density of states in the absence of spin orbit interaction. The superscripts $R$ and $A$ stand for retarded and advanced Green functions, respectively.
The first equation represents the ladder resummation for an effective vertex $\tilde \gamma_x$, which is defined by the second equation. 
Notice that the first term in the expression for $\tilde \gamma_x$ is the bare vertex of Eq.~(\ref{a_4c}). 
In the purely two-dimensional Rashba model with no $z$-dependence, one sees that the second term on the right hand side of the second equation cancels the first. This is the famous vertex cancellation. To see this explicitly one must project the above equation into the spin states $|\lambda \rangle$, the projection over the plane wave states already being done. 

To extend the treatment to the present case, the projection must be made over the states $|n \lambda \rangle$.
Within the approximation of disorder with no $z$-dependence of the impurity potential, the vertex equations are not changed. The second of the Eqs.~ (\ref{a_7}) becomes
\ber
\label{a_8}
\langle n\kv \lambda |{\tilde \gamma}_x|n'\kv \lambda' \rangle&=&-\langle n\kv \lambda |\alpha \Theta (z)\sigma^y|n' \kv \lambda' \rangle\nn\\&+&
\frac{1}{2\pi N_0 \tau}\sum_{n_1 \kv'\lambda_1} \langle n\kv \lambda  |n_1\kv' \lambda_1 \rangle G^R_{n_1 \kv'\lambda_1} \ 2 k_x' \ G^A_{n_1\kv'\lambda_1}\langle n_1 \kv' \lambda_1   |n' \kv \lambda' \rangle.
\eer
The matrix elements $\langle n \kv \lambda  |n_1 \kv' \lambda_1 \rangle$ and $\langle n_1 \kv' \lambda_1   |n' \kv \lambda' \rangle$ are those of the impurity potential. Explicitly we have
\begin{eqnarray}
\langle n \kv \lambda  |n_1 \kv' \lambda_1 \rangle&=& \frac{1}{2}
\langle f_{n\kv\lambda}|f_{n_1\kv'\lambda_1}\rangle \left( 1+\lambda \lambda_1 e^{{\rm i}(\theta_{\kv'}-\theta_{\kv})}\right)\\
\langle n_1 \kv' \lambda_1   |n' \kv \lambda' \rangle&=&
\frac{1}{2}
\langle f_{n_1\kv'\lambda_1}|f_{n'\kv\lambda'}\rangle \left( 1+\lambda' \lambda_1 e^{-{\rm i}(\theta_{\kv'}-\theta_{\kv})}\right).
\end{eqnarray}
By observing that $k_x'=k' \cos \theta_{\kv'}$, one can perform the integration over the direction of $\kv'$ in  Eq.(\ref{a_8})
\begin{equation}
\label{a_8a}\frac{1}{4}
\int_0^{2\pi} \frac{{\rm d}\theta_{\kv'}}{2\pi}
\left( 1+\lambda \lambda_1 e^{{\rm i}(\theta_{\kv'}-\theta_{\kv})}\right)
\cos (\theta_{\kv'})
\left( 1+\lambda' \lambda_1 e^{-{\rm i}(\theta_{\kv'}-\theta_{\kv})}\right)=\frac{\lambda_1}{8}\left( \lambda e^{-{\rm i}\theta_{\kv}}+
\lambda'  e^{{\rm i}\theta_{\kv}}\right).
\end{equation}
We now rewrite Eq.(\ref{a_8}) for the matrix elements necessary
for the evaluation of the spin Hall conductivity.  By using Eq.(\ref{a_4c})
\ber
\label{a_8b}
\langle 0\kv \lambda |{\tilde \gamma}_x|n\kv \overline{\lambda} \rangle&=&
-{\rm i}  \frac{\sin (\theta_{\kv})}{2k}(E_{0\kv\lambda} 
-E_{n\kv\overline{\lambda}})\langle f_{0\kv\lambda}|f_{n\kv\overline{\lambda}}\rangle
\nn\\
&-&{\rm i}
\frac{\sin (\theta_{\kv})\lambda }{8\pi N_0 \tau}\sum_{n_1 \kv'\lambda_1}\lambda_1  
\langle f_{0\kv\lambda}|f_{n_1\kv'\lambda_1}\rangle 
\langle f_{n_1\kv'\lambda_1}|f_{n\kv\overline{\lambda}}\rangle
 G^R_{n_1 \kv'\lambda_1} \ 2 k' \ G^A_{n_1\kv'\lambda_1}.
\eer
For weak disorder, we may take the limit $\tau\rightarrow\infty$ and perform the integral over $\kv'$ to get
\ber
\label{a_8bb}
\langle 0\kv \lambda |{\tilde \gamma}_x|n\kv \overline{\lambda} \rangle&=&
-{\rm i}  \frac{\sin (\theta_{\kv})}{2k}(E_{0\kv\lambda} 
-E_{n\kv\overline{\lambda}})\langle f_{0\kv\lambda}|f_{n\kv\overline{\lambda}}\rangle
\nn\\
&-&{\rm i}
\frac{\sin (\theta_{\kv})\lambda }{2 N_0 }\sum_{n_1 \lambda_1}\lambda_1  
\langle f_{0\kv\lambda}|f_{n_1k_{Fn_1\lambda_1}\lambda_1}\rangle 
\langle f_{n_1k_{Fn_1\lambda_1}\lambda_1}|f_{n\kv \overline{\lambda}}\rangle
  \  k_{Fn_1\lambda_1} \ N_{n_1\lambda_1},
\eer
$k_{Fn_1\lambda_1}$ and $N_{n_1\lambda_1}$ being the Fermi momentum and the density of states at the chemical potential for the $n_1$-th subband
with helicity $\lambda_1$.
Let us consider first the intra-band matrix elements, $n=0$, so that Eq.(\ref{a_8bb})  becomes
\ber
\label{a_8c}
\langle 0\kv \lambda |{\tilde \gamma}_x|0\kv \overline{\lambda} \rangle&=&
-{\rm i}  \frac{\sin (\theta_{\kv})}{2k}(E_{0\kv\lambda} 
-E_{0\kv\overline{\lambda}})\langle f_{0\kv\lambda}|f_{0\kv\overline{\lambda}}\rangle
\nn\\
&-&{\rm i}
\frac{\sin (\theta_{\kv})\lambda }{2 N_0 }\sum_{n_1 \lambda_1}\lambda_1  
\langle f_{0\kv\lambda}|f_{n_1k_{Fn_1\lambda_1}\lambda_1}\rangle 
\langle f_{n_1k_{Fn_1\lambda_1}\lambda_1}|f_{0\kv \overline{\lambda}}\rangle
  \  k_{Fn_1\lambda_1} \ N_{n_1\lambda_1},
\eer
Now we show how this equation can be evaluated in a small $\alpha$ expansion up to the third order.  We begin by observing that by performing 
the sum over $\lambda_1$, always one of the overlap factors yields a Kronecker $\delta_{n_1,0}$ because of the orthonormality of the wave functions $f_{n\kv\lambda}$. Furthermore, because
of Eqs.(\ref{Overlap-nn2}), we can replace $k_{Fn_1\lambda_1}$ with $\kv$ with an accuracy
$\alpha^3$. In this way the overlap factor between states of the same helicity
 can be approximated with unity, while the other is common to the bare part of the vertex
 (the first term on the right hand side of Eq.(\ref{a_8c})).
 Hence we are  reduced to evaluating
\ber
\label{a_8cc}
\frac{\langle 0\kv \lambda |{\tilde \gamma}_x|0\kv \overline{\lambda} \rangle}{
\langle f_{0\kv\lambda}|f_{0\kv\overline{\lambda}}\rangle}=
-{\rm i}  \frac{\sin (\theta_{\kv})}{2k}(E_{0\kv\lambda} 
-E_{0\kv\overline{\lambda}})
-{\rm i}
\frac{\sin (\theta_{\kv})\lambda }{2 N_0 }\sum_{ \lambda_1}\lambda_1  
  \  k_{Fn_1\lambda_1} \ N_{n_1\lambda_1}.
\eer
%
 The densities of states at the Fermi level in the lowest subband are evaluated from the formula
 \begin{equation}
 \label{soe_6}
 \frac{N_\lambda}{N_0}=
 \frac{2 k_\lambda}{|\nabla_k E_{0k\lambda}|_{k_\lambda}}\,,
 \end{equation}
 where $k_\lambda$ is the solution of the equation $E_{0k\lambda}=E_{F}$.
By using the equations~(\ref{EnergyExpansion}-\ref{soe_14}) for the energy and  (\ref{soe_2}-\ref{soe_5}) for the Fermi wave vectors, we obtain 
 \begin{equation}
 \label{soe_7}
 \frac{N_\lambda}{N_0}=1-\lambda \alpha\frac{e_1}{2k_F}-\alpha^2 e_2
 +\lambda \alpha^3 \left[ \frac{e_1e_2}{4k_F}-\frac{3e_3k_F}{2}+\frac{e_1^3}{16k_F^3}\right].
 \end{equation}
 The quantity relevant for the vertex correction is
 \begin{equation}
 \label{soe_8}
  \frac{k_\lambda N_\lambda}{2N_0}=\frac{k_F}{2}
  \left\{ 1-\lambda \alpha\frac{e_1}{k_F}  +\alpha^2\frac{3}{k_F}\left[\frac{e_1^2}{8k_F} -\frac{e_2k_F}{2}\right] +\lambda \alpha^3\frac{4}{k_F}
  \left[ \frac{e_1e_2}{2}-\frac{e_3k_F^2}{2}\right] \right\}.
 \end{equation}
 The vertex correction is then
 \begin{equation}
 \label{soe_9}
 \frac{k_\lambda N_\lambda}{2N_0}-
 \frac{k_{\bar\lambda} N_{\bar\lambda}}{2N_0}=- \lambda\alpha e_1+
 2 \lambda \alpha^3 \left( e_1e_2 -e_3 k_F^2\right),
 \end{equation}
 while the bare vertex is
  \begin{equation}
 \label{soe_10}
 \frac{1}{2k}\left( E_{0 k \lambda} -E_{0 k \bar\lambda}\right)=\lambda \alpha e_1 +\lambda \alpha^3 e_3 k^2\,,
 \end{equation}
 where, in the last term, $k$ can be replaced by $k_F$ at the required level of accuracy. 
 By summing the above two expressions, the effective vertex reads
 \begin{equation}
 \label{soe_11}
 \langle 0\kv \lambda |{\tilde \gamma}_x|0\kv \overline{\lambda} \rangle=
-{\rm i} \lambda  \sin (\theta_{\kv})\langle f_{0\kv\lambda}|f_{0\kv\overline{\lambda}}\rangle
 \alpha^3 \left( 2e_1e_2 -e_3 k_F^2\right).
 \end{equation}
 The above expression must replace Eq.(\ref{a_4c}) when evaluating the intra-band spin Hall conductivity and one can also safely neglect the ladder resummation in the first equation of (\ref{a_7}) due to the $\tau\rightarrow\infty$ limit. Hence,
to the accuracy we are working it is enough to multiply the {\sl bare bubble} intra-band
spin Hall conductivity of Eq.(\ref{SHC-Intra1}) by the factor $-\alpha^2 \left( 2e_1e_2 -e_3 k_F^2\right)$.
Notice that in obtaining  Eq.(\ref{SHC-Intra1}), the momentum integral must be evaluated with accuracy $\alpha^3$ in order to keep the corrections up to order $\alpha^2$ in the SHC. The presence of the corrected vertex, which is already
of the order $\alpha^2$ smaller than the bare one, allows the evaluation of the integral with accuracy $\alpha$.
Notice also  that in the Rashba limit, $r\rightarrow \infty$, the coefficients $p_{0n}$ vanish, and hence $e_2$ and $e_3$ both vanish,  yielding the vertex cancellation as expected.
 The intra-band spin Hall conductivity of Eq.(\ref{SHC-Intra1}) must then be replaced by 
 \begin{equation}
 \label{SHC-Intra2}
 \left[ \sigma^{intra}_{SH}\right]^z_{yx}=-\frac{e}{8\pi}\alpha^2
 \left\{ 2\sum_{n\neq0}\frac{|p_{0n}|^2}{\epsilon_n -\epsilon_0}+\frac{k_F^2}{p_{00}}
\left[ \sum_{n,m\neq0}\frac{p_{0n}p_{nm}p_{m0}}{(\epsilon_n-\epsilon_0)(\epsilon_m -\epsilon_0)}-\sum_{n\neq0} \frac{|p_{0n}|^2}{(\epsilon_n-\epsilon_0)^2}\right]\right\},
 \end{equation}
 which is accurate up to terms of order $\alpha^3$.  This can also be written as
 \begin{equation}
 \label{SHC-Intra3}
 \left[ \sigma^{intra}_{SH}\right]^z_{yx}=-\frac{e}{4\pi}\left\{\left(1-\frac{m}{m^*}\right) +\frac{\alpha^2 k_F^2}{2 p_{00}}
\left[\sum_{n,m\neq0}\frac{p_{0n}p_{nm}p_{m0}}{(\epsilon_n-\epsilon_0)(\epsilon_m -\epsilon_0)}-\sum_{n\neq0} \frac{|p_{0n}|^2}{(\epsilon_n-\epsilon_0)^2}\right]\right\},
 \end{equation}
\indent The numerically calculated intra-subband contribution to the spin Hall conductivity from Eq.~(\ref{SHC-Intra2})  is plotted (in units of $\frac{e\alpha^2}{\pi}$ ) in Fig.4. We note that it is present for all values of r, but it vanishes in the Rashba limit ($r \rightarrow\infty$) because $p_{nm}$ vanishes in that limit. 

 \begin{figure}
\begin{center}
\includegraphics[width=3.0in]{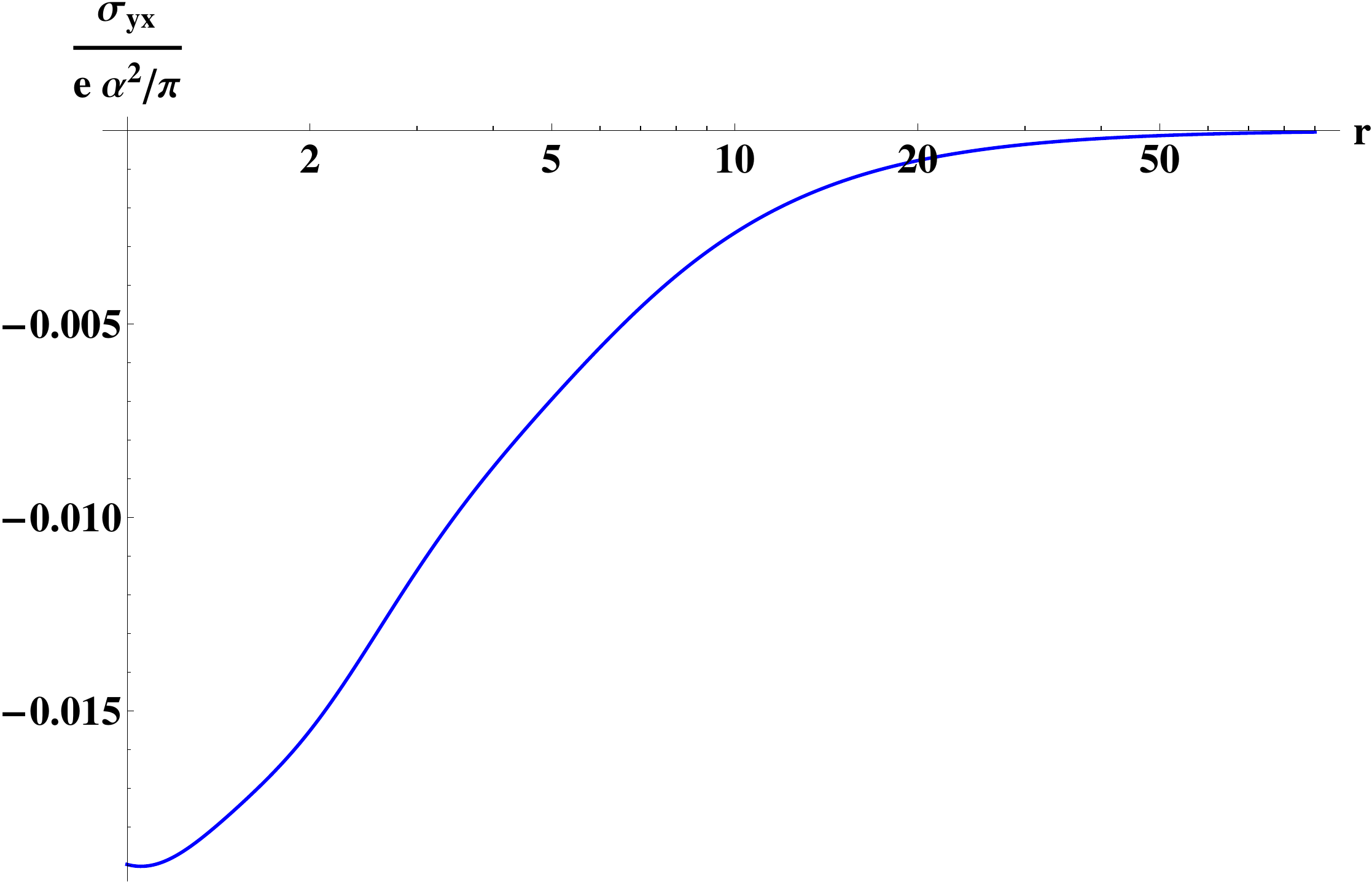}
\caption{Intrasubband contribution to  the spin Hall conductivity in units of $\frac{e \alpha^2}{\pi}$ for $k_F=1$(see Eq.~(\ref{InterSubband})).   Data points have been connected to enhance visibility of the trend.  Notice the logarithmic scale for $r$.} 
\end{center}
\label{Fig4}
\end{figure}
	
Vertex corrections, in principle, are present also for the inter-band matrix elements controlling the inter-band spin Hall conductivity.
In the case of the matrix elements between the occupied and unoccupied sub-bands, with accuracy of order $\alpha$, Eq.(\ref{a_8bb}) becomes
\begin{equation}
\label{a_8g}
\langle 0\kv \lambda |{\tilde \gamma}_x|n\kv \overline{\lambda} \rangle=
-{\rm i}  \sin (\theta_{\kv})\lambda \alpha p_{0n}
-{\rm i}
\frac{\sin (\theta_{\kv})\lambda }{2N_0}\sum_{n_1 \lambda_1}\lambda_1  
\langle f_{0k_F\lambda}|f_{n_1k_F\lambda_1}\rangle 
\langle f_{n_1k_F\lambda_1}|f_{nk_F\overline{\lambda}}\rangle
k_{F n_1\lambda_1} N_{n_1\lambda_1}  .
\end{equation}
By performing the sum over $\lambda_1$ and using again the orthonormality of the wave functions $f_{n\kv\lambda}$, one obtains
\begin{equation}
\label{a_8h}
\langle 0\kv \lambda |{\tilde \gamma}_x|n\kv \overline{\lambda} \rangle=
-{\rm i}  \sin (\theta_{\kv})\lambda \alpha p_{0n}
-{\rm i}
\sin (\theta_{\kv}) \alpha p_{0n}\frac{p_{00}+p_{nn}}{ \epsilon_n-\epsilon_0}.
\end{equation}
We see that the first vertex correction for the inter-band matrix elements contains as a denominator the energy separation.  For large enough separation, this correction can be neglected.  It is interesting to note that the energy separation in the denominator appears because of the overlap factors due to the impurity potential scattering. The physical origin of these processes is the following. Upon scattering from an impurity an electron can make a transition to an unoccupied subband, where afterwards changes its spin direction by precessing in the Rashba field. Clearly these processes cost energy and yield a small correction to the spin-dependent inter-band matrix elements of the current vertex.
As a result, Eq.(\ref{InterSubband}) for the inter-band spin Hall conductivity is not changed much, either qualitatively or quantitatively, by the vertex corrections.

\begin{figure}
	\begin{center}
	\includegraphics[width=3.0in]{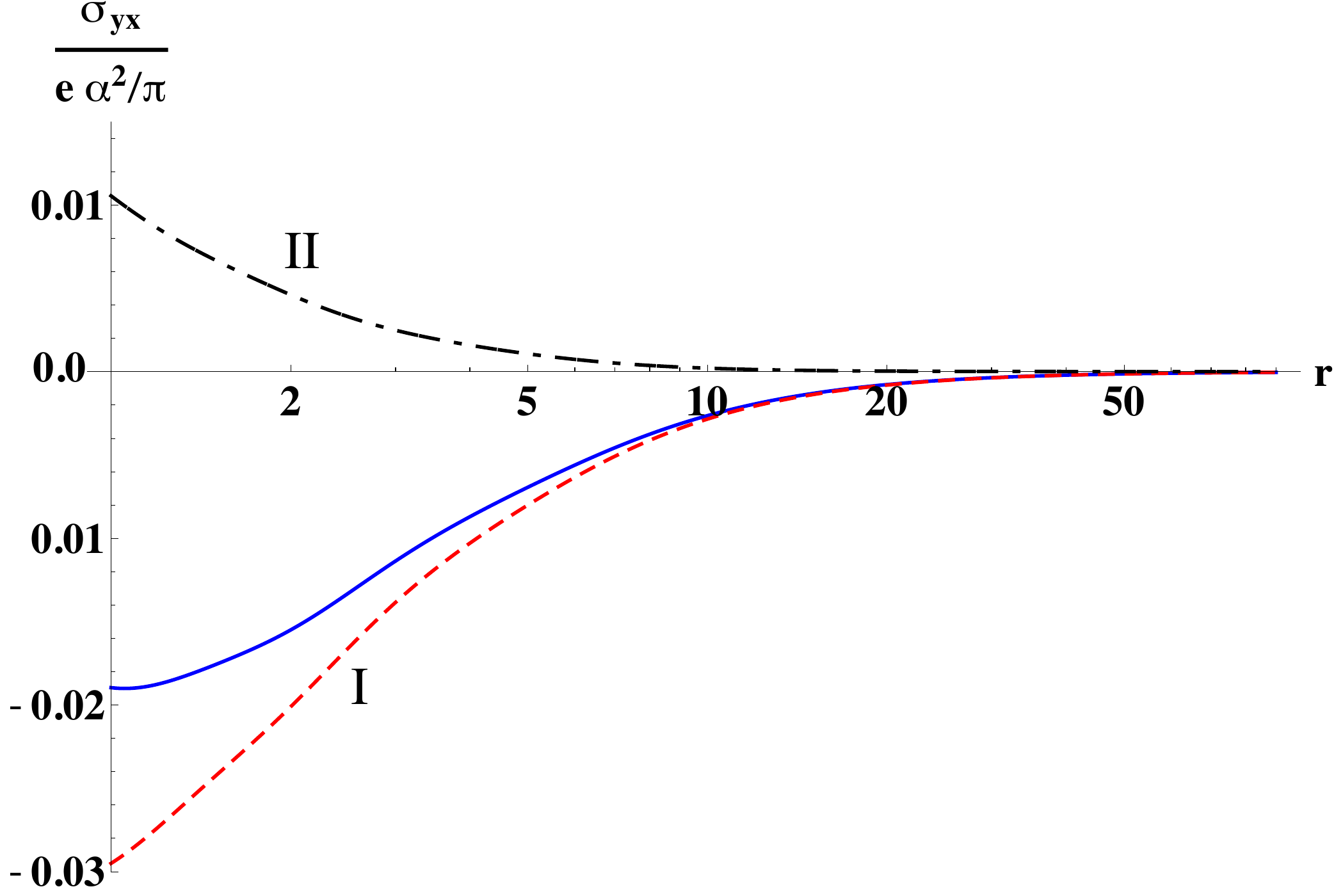}
	\caption{Comparison of the first and the second term in the curly brackets of Eq.~(\ref{SHC-Intra3}) for the intra-band spin Hall conductivity at $k_F=1$.  First term, denoted by I,  is in red (dashed line), second term, denoted by II, is in black (dash-dotted line); their sum is in blue (solid line).  Notice that the two terms have opposite signs and I dominates.} 
	\end{center}
	\label{Fig5}
	\end{figure}

Eqs.~(\ref{InterSubband}) and (\ref{SHC-Intra3})   are the main results of this paper, showing a spin Hall conductivity of order $\alpha^2$.
Notice that, in the low-density limit $k_F \to 0$ only the intra-band term survives,  reducing to the simple form
 \be\label{SHC4}
 \left[ \sigma_{SH}\right]^z_{yx}=-\frac{e}{4\pi}\left(1-\frac{m}{m^*}\right)\,, 
 \ee
 which is entirely controlled by the spin-orbit-induced effective mass.
 Fig. 5 examines the relative importance of the two terms in the curly brackets of Eq.~(\ref{SHC-Intra3})  at $k_F=1$.   The effective mass term (first term) is plotted in red, the second term is plotted in black, and their sum is in blue (this coincides with the result plotted in Fig. 4). 
 We see that   the two terms in the curly brackets of Eq.~(\ref{SHC-Intra3})  have opposite signs, but the first term dominates even at $k_F$ as large as $1$.

\section{Conclusion}
We have developed a simple model for the 2DEG that exists at the interface between two oxides.  Neglecting  band structure effects and including only an asymmetric wedge-shaped potential that binds the electrons to the interface, and the spin-orbit interaction associated with it, we have calculated the intrinsic spin Hall conductivity in the high-mobility limit.
After a careful consideration of vertex corrections to the single-bubble result, we have found that the intrinsic SHC is finite and, in the low-carrier density limit, has the simple form of Eq.~(\ref{SHC4}), which crucially depends on the mass renormalization associated with the k-dependence of the wave functions in the $z$ direction.  This effect, which is of order $\alpha^2$, vanishes in the standard Rashba model, which is the $r \to \infty$ limit of the present model.
Finally, the numerical value of the SHC that we calculate is of the order of $\frac{e}{8 \pi} 0.08 \alpha^2$.   This should be compared with the values measured in the two-dimensional electron gas in GaAs,\cite{Sih05} which can be expressed as $\frac{e}{8 \pi} 0.05 \alpha^2$.  Thus our values  are of a magnitude comparable with those reported in GaAs  if $\alpha$ is of order $1$ as reported in the literature.\cite{Caviglia10}

Our model clearly does {\it not} include the spin-orbit interaction that is built into the two-dimensional band structure of the interfacial electrons, arising from the spin-orbit interaction with the atomic cores.  The latter can be extracted from a tight-binding model calculation for SrTiO$_3$, taking into account the fact that the interfacial electrons live almost entirely on the STO side of the LAO/STO interface.\cite{Ismail-Beigi2010}  From the form of the in-plane wave function in the relevant conduction band (arising from Ti d-orbitals split by crystal fields and  spin-orbit interaction with the atomic cores) one can extract an effective spin-orbit coupled hamiltonian,~\cite{Khalsa13} and the intrinsic spin Hall conductivity can be calculated.  A comparison between the SHC calculated in this paper and that obtained from a tight-binding calculation of the spin-orbit coupled band structure will be presented in a forthcoming paper.

 \section{Acknowledgement}
  We acknowledge  support from ARO Grant No. W911NF-08-1-0317 and from
 EU through Grant No. PITN-GA-2009-234970.  We thank for discussions M. Grilli and S. Caprara.

\bibliography{refs_SJ-1}
\end{document}